\documentclass[aps,prl,reprint,superscriptaddress,nofootinbib,nobibnotes,preprintnumbers]{revtex4-2}

\usepackage{amsmath,amssymb}
\usepackage{graphicx}
\usepackage{placeins}
\usepackage{xcolor}
\definecolor{lightblue}{RGB}{5,171,240}

\usepackage[colorlinks=true,allcolors=blue]{hyperref}

\newif\ifsupplementaltoc
\makeatletter
\newcommand{\supplementalcontents}{%
  \begingroup
  \supplementaltocfalse
  \let\supplement@contentsline\contentsline
  \renewcommand{\contentsline}[4]{%
    \ifsupplementaltoc
      \supplement@contentsline{##1}{##2}{##3}{##4}%
    \fi
  }%
  \let\tocdepth@munge\relax
  \let\tocdepth@restore\relax
  \let\addcontentsline\@gobblethree
  \tableofcontents
  \endgroup
}
\makeatother

\begin{document}

\title{TeV Gamma-Ray Transparency Bounds on Neutrinos and Axions}

\makeatletter
\author{}
\affiliation{School of Fundamental Physics and Mathematical
Sciences, Hangzhou Institute for Advanced Study, UCAS, Hangzhou 310024, China}
\move@AF\move@AUAF
\let\@AAC@list\@empty
\let\@AFG@list\@empty
\makeatother

\author{Shengyi Liu}
\email{sliufp@connect.ust.hk}
\affiliation{Department of Physics and Jockey Club Institute for Advanced Study,
The Hong Kong University of Science and Technology, Hong Kong S.A.R., China}

\author{Hyeonseok Seong}
\email{hseong0627@gmail.com}
\affiliation{State Key Laboratory of Dark Matter Physics, Tsung-Dao Lee Institute, Shanghai Jiao Tong University, Shanghai 200240, China} 
\affiliation{Key Laboratory for Particle Astrophysics and Cosmology (MOE) \& Shanghai Key Laboratory for Particle Physics and Cosmology, Shanghai Jiao Tong University, Shanghai 200240, China}
\affiliation{Department of Physics and Jockey Club Institute for Advanced Study,
The Hong Kong University of Science and Technology, Hong Kong S.A.R., China}
\affiliation{Deutsches Elektronen-Synchrotron DESY, Notkestr.~85,
22607 Hamburg, Germany}

\author{Chen Sun}
\email{chen.sun@ucas.ac.cn}
\affiliation{School of Fundamental Physics and Mathematical
Sciences, Hangzhou Institute for Advanced Study, UCAS, Hangzhou 310024, China}
\affiliation{International Centre for Theoretical Physics Asia-Pacific, Beijing / Hangzhou, China}
\affiliation{Abdus Salam International Centre for Theoretical Physics, Strada Costiera 11, 34151, Trieste, Italy}

\author{Seokhoon Yun}
\email{seokhoon.yun@knu.ac.kr}
\affiliation{Department of Physics, Kyungpook National University,
Daegu 41566, Korea}
\affiliation{Particle Theory and Cosmology Group, Center for Theoretical Physics
of the Universe, Institute for Basic Science, Daejeon 34126, Korea}

\date{\today}

\begin{abstract}
Multi-TeV gamma rays propagating over cosmological distances measure the extragalactic background light (EBL) through attenuation by electron--positron pair production. This is particularly valuable in the infrared, where the EBL is difficult to measure directly. Since the growing sample of observed multi-TeV photons provides tight constraints on the EBL photon intensity, this, in turn, constrains new particles that radiatively decay into meV--eV photons.  We set the first TeV-transparency bounds on active-neutrino electromagnetic transition moments, obtain new constraints on radiatively decaying sterile neutrinos, and derive competitive limits on the axion--photon coupling.
\end{abstract}


\maketitle

\section{Introduction}

The extragalactic background light (EBL) is the accumulated optical and
infrared radiation produced over the history of galaxies.  Unlike the cosmic
microwave background, its main contribution is purely astrophysical.  
Predicting it requires a synthesis of stellar emission, dust
reprocessing, galaxy evolution, and the cosmic star-formation history
\cite{Hill:2018trh}.  Consequently, the photon density relevant for the
propagation of very-high-energy gamma rays is difficult to compute.

Observationally, the EBL can be determined in three complementary ways.
Summed light from resolved galaxies provides a lower bound
\cite{Driver:2016}, subject to the completeness of the galaxy catalog.
Direct photometry measures the total intensity, but only after bright
foregrounds have been subtracted \cite{Postman:2024}.  Finally, attenuation
of multi-TeV gamma rays measures the EBL along cosmological lines of sight.
Joint fits can reconstruct the EBL spectrum together with the intrinsic
blazar spectra \cite{Yuan:2012EBL}, an approach also explored for LHAASO
blazar observations \cite{Wang:2022LHAASOExtragalactic}.
A recent analysis of more than 260 archival TeV spectra reconstructed the EBL from
$0.18$ to $120\,\mu\mathrm{m}$ and found no significant diffuse excess above
integrated galaxy light \cite{Greaux:2024}.  This agreement makes it possible
to constrain new physics that generates an additional cosmological photon
population, which is the focus of this Letter.

Throughout, $E_{\rm UV}$ denotes the gamma-ray energy observed today, while
$E_{\rm IR}$ denotes the EBL photon's local energy at the interaction
redshift $z$.  The gamma ray's local energy is therefore
$(1+z)E_{\rm UV}$.  The Breit--Wheeler process
$\gamma_{\rm UV}\gamma_{\rm IR}\to e^+e^-$ is allowed when
\begin{equation}
  2(1+z)E_{\rm UV}E_{\rm IR}(1-\cos\theta)\geq 4m_e^2,
  \label{eq:pair-threshold}
\end{equation}
where $\theta$ is the collision angle.  The pair-production {cross section and threshold kinematics} therefore
makes photons above $\mathcal{O}(10)\,\mathrm{TeV}$ especially sensitive to the
part of the EBL above $0.01~\mathrm{eV}$.  For a source at redshift $z_s$,
let $\Phi_{\rm int}(E_{\rm UV})$ be the flux that would reach us without EBL
absorption.  The observed flux is then
$\Phi_{\rm obs}(E_{\rm UV})=\Phi_{\rm int}(E_{\rm UV})
e^{-\tau(E_{\rm UV},z_s)}$.  An observed multi-TeV spectrum therefore constrains
the optical depth and hence a cross-section-weighted photon column density.
The intrinsic source spectrum remains a nuisance quantity and is treated
explicitly below.

GRB~221009A provides a particularly instructive example.  At redshift
$z=0.151$, LHAASO detected more than 140 photons above $3\,\mathrm{TeV}$ and
emission extending beyond $10\,\mathrm{TeV}$ \cite{LHAASO:2023}.  Such photons
traverse a region in which the predicted optical depth varies substantially
among EBL prescriptions \cite{Baktash:2022gnf}.  Accordingly, how anomalous the
event appears is inseparable from the assumed EBL model and from the inferred
intrinsic burst spectrum.  Photon--axion conversion has been tested with
blazar spectra under different EBL assumptions \cite{Guo:2021ALPEBL} and
with the LHAASO spectrum of this burst \cite{Gao:2024GRBALP}.
We therefore do not use GRB~221009A as motivation
for a mechanism that evades standard attenuation.  Instead, we take its
observed high-energy spectrum as the latest empirical update of the EBL: the
survival of the photons limits the number density of infrared targets that can
be present along the line of sight.

This perspective turns TeV transparency into a probe of particles that inject
soft photons into the Universe~\cite{Bernal:2023baf}.  Radiative decays such as
neutrino dipole transitions
$(\nu_i\to\nu_j\gamma)$,
sterile neutrinos $(\nu_s\to\nu\gamma)$, and
axions $(a\to\gamma\gamma)$~\cite{Bernal:2023baf} generate
diffuse radiation in precisely the energy range
selected by multi-TeV gamma rays.  Since this contribution is non-negative, it
can only increase the pair-production optical depth.  The observed
transparency thus places an upper bound on the additional photon density
$\delta n$ and, in turn, on the relevant decay rates and couplings.
For axions, infrared-background anisotropies have been used to constrain
the decay contribution \cite{Gong:2016AxionEBL}, and additional gamma-ray
absorption from the resulting EBL photons has been considered in LHAASO
sensitivity forecasts \cite{Bi:2022LHAASONewPhysics}.

In this Letter we combine the EBL reconstruction based on STeVECat
\cite{Greaux:2023STeVECat,Greaux:2024} with the LHAASO spectrum of
GRB~221009A \cite{LHAASO:2023} to obtain the strongest constraint to date on
the EBL photon density in the energy range relevant for multi-TeV attenuation.
Interpreting any remaining EBL intensity as a possible new-physics photon
contribution, $\delta n$, we 
set the first
TeV-transparency bounds on active-neutrino electromagnetic transition moments surpassing CMB constraint, obtain new constraints on radiatively decaying sterile neutrinos, and derive the competitive limit on the axion--photon coupling {on par with some of the strongest constraints in certain mass window.}


\section{Attenuation of TeV photons}

In this and subsequent sections, we use natural units with
$\hbar=c=k_{\rm B}=1$.
The quantity measured by gamma-ray attenuation is the pair-production optical
depth, rather than an unweighted photon column density.  For a gamma ray
observed with energy $E_{\rm UV}$ from a source at redshift $z_s$, propagation
through an isotropic photon background gives
\begin{align}
 \tau_{\gamma}(E_{\rm UV},z_s)
 ={}& \int_0^{z_s}\!dz\,
 \frac{1}{(1+z)H(z)}
 \int_{-1}^{1}\!d\mu\,\frac{1-\mu}{2}
 \nonumber\\[-2pt]
 &\times \int_{E_{\rm IR}^{\rm th}}^\infty\!dE_{\rm IR}\,
 \frac{dn_\gamma}{dE_{\rm IR}}(E_{\rm IR},z)
 \sigma_{\gamma\gamma}(s),
 \label{eq:optical-depth}
\end{align}
where $\mu\equiv\cos\theta$, with $\theta$ the angle between the two photon
momenta,
$s=2E_{\rm UV}(1+z)E_{\rm IR}(1-\mu)$, and
$E_{\rm IR}^{\rm th}=2m_e^2/[E_{\rm UV}(1+z)(1-\mu)]$.
Here $E_{\rm IR}$ is the local EBL photon energy, and
$dn_\gamma/dE_{\rm IR}$ is the proper spectral number density, both evaluated
at redshift $z$.  The Breit--Wheeler cross section \cite{Gould:1967} vanishes below
threshold and peaks not far above it, {making the attenuation energy range moderately broad around $E_{\rm IR}$.}
Explicitly,

\begin{align}
 \sigma_{\gamma\gamma}(s)
 ={}& \frac{3\sigma_{\rm T}}{16}(1-\beta_s^2)
 \left[
  (3-\beta_s^4)\ln\!\left(\frac{1+\beta_s}{1-\beta_s}\right)
  \right.\nonumber\\[-2pt]
 &\left.{}-2\beta_s(2-\beta_s^2)\right],
 \label{eq:breit-wheeler}
\end{align}
for $s\geq 4m_e^2$, where
$\beta_s=\sqrt{1-4m_e^2/s}$ and $\sigma_{\rm T}$ is the Thomson cross section;
below threshold, $\sigma_{\gamma\gamma}=0$.  


In the analysis, we use the EBL reconstructed from more than 260 archival TeV spectra
\cite{Greaux:2024}. We also take into account the LHAASO observation of GRB~221009A \cite{LHAASO:2023}, which extends the sensitivity to the EBL at the infrared energy. 
To constrain the photon density injected by new physics, we
decompose the photon field as
\begin{equation}
 \frac{dn_\gamma}{dE_{\rm IR}}
 =\frac{dn_\gamma^{\rm gal}}{dE_{\rm IR}}
 +\frac{dn_\gamma^{\rm NP}}{dE_{\rm IR}},
 \label{eq:ebl-decomposition}
\end{equation}
Here $n_\gamma$ is the total proper EBL photon number density,
$n_\gamma^{\rm gal}$ is the standard astrophysical component generated by
stellar emission and dust reprocessing over the cosmic star-formation
history, and $n_\gamma^{\rm NP}$ is an additional diffuse component produced by new
physics.  Since Eq.~\eqref{eq:optical-depth} is linear in the target photon
density, the same decomposition applies to the optical depth,
$\tau_\gamma=\tau_\gamma^{\rm gal}+\tau_\gamma^{\rm NP}$.

\begin{figure}[t]
  \centering
  \includegraphics[width=\columnwidth]{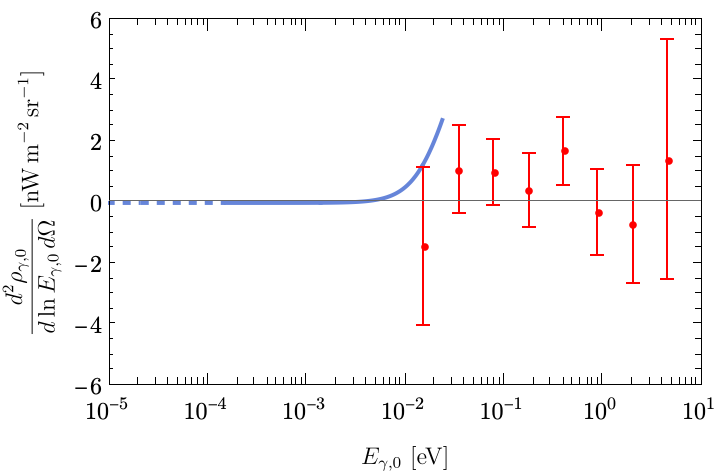}
  \caption{Present-day residual EBL (red), obtained by subtracting galaxy-count
  IGL from the gamma-ray reconstruction~\cite{Driver:2016,Greaux:2024}.
  Blue: active-neutrino decay with $m_{\rm i}=0.051\,\mathrm{eV}$,
  $m_{\rm f}=0.013\,\mathrm{eV}$, and $\mu_{\rm eff}=10^{-9}\mu_{\rm B}$.
  Contributions exceeding the allowed residual predict excessive TeV attenuation.
  }
  \label{fig:ebl-excess-new-physics}
\end{figure}

Integrated light from resolved galaxies provides a measured floor for
$n_\gamma^{\rm gal}$, while the new-physics contribution must be non-negative.
We conservatively assign all EBL intensity allowed above that floor to
$n_\gamma^{\rm NP}$ and require the resulting optical depth to remain compatible
with the gamma-ray data.  Any unresolved astrophysical emission occupies part
of the same residual and would therefore strengthen the bounds derived below.
Figure~\ref{fig:ebl-excess-new-physics} illustrates this constraint on a
diffuse photon population generated by new physics.

\section{Intrinsic Gamma-Ray Spectra of Extragalactic Sources}

Extracting the EBL optical depth requires separating propagation from the
emission of the source:
\begin{align}
    \Phi_{\rm obs}(E,z)&=e^{-\tau_\gamma(E,z)}\Phi_{\rm int}(E),
\end{align}
{where the intrinsic spectrum $\Phi_{\rm int}(E)$ can be parametrized as \cite{Greaux:2024}, for example, }
\begin{equation}
\begin{aligned}
 \Phi_{\rm int}(E)&=\Phi_0\exp\!\left[
 \eta-\Gamma\ln\frac{E}{E_0}
 -\beta\ln^2\!\frac{E}{E_0}-\lambda E\right].
\end{aligned}
 \label{eq:intrinsic-gamma-ray-spectrum}
\end{equation}
The intrinsic spectrum is a flexible phenomenological description of a smooth source
continuum.  It combines a log-parabolic spectrum with an exponential cutoff.
Here $E_0$ is a fixed pivot energy and $\Phi_0$ is the corresponding reference
flux, while
the logarithmic normalization $\eta$, index $\Gamma$, curvature $\beta$, and
inverse cutoff $\lambda$ (with $\lambda=E_{\rm cut}^{-1}$ for an ordinary
cutoff) are fitted separately for every observation, to account for the intrinsic variation of the spectra.  The EBL reconstruction
marginalizes over these source parameters rather than fixing an intrinsic
spectrum.

The source shape is anchored with gamma-ray data below the strongly absorbed
band.  The STeVECat analysis combines 268 VHE spectra of 45 sources, measured
mainly by H.E.S.S., MAGIC, and VERITAS, with contemporaneous Fermi-LAT spectra
whenever available.  For 64 observations, the $0.1$--$100\,\mathrm{GeV}$ LAT
spectrum---where EBL attenuation is negligible for the redshifts considered---
provides a correlated prior on $(\Gamma,\beta)$.  Its normalization is left
free because blazars vary, sometimes on timescales shorter than an observing
window; LAT therefore anchors the shape, not the TeV flux level.  Spectra
without contemporaneous LAT coverage are assigned weakly informative shape
priors and are marginalized in the same way.

Intrinsic curvature and EBL attenuation are thus not assumed to be exactly
orthogonal: a source cutoff can mimic some additional softening.  The converse
degeneracy relevant to an upper limit on $\tau_\gamma$ is much weaker.  Hiding
a large optical depth would require the emitted spectrum to compensate the
factor $e^{-\tau_\gamma}$ through an exponential hardening,
$\Phi_{\rm int}\propto e^{+\tau_\gamma}$, at the highest energies.  Such an
upturn is not expected from standard particle-acceleration and radiative-loss
physics.  In addition, the EBL imprint is shared by many sources and scales
with redshift, whereas the intrinsic parameters vary independently.  The
external low-energy anchor and the joint source sample therefore leave only a
limited degeneracy between intrinsic shape and optical depth.

\section{Constraints on new physics}

We now translate the allowed non-astrophysical component of the EBL, $n^{\rm NP}_\gamma$, into limits on
radiative decays.  Consider a homogeneous population of nonrelativistic
particles $X$, each of which emits $N_\gamma$ photons with rest-frame energy
$E_*$ and radiative width $\Gamma_X$.  If the decay photons propagate freely,
cosmological redshifting broadens the monochromatic emission into the
present-day spectrum.  Evaluating $E_{\rm IR}$ and the photon density at
$z=0$ gives
\begin{equation}
 \frac{dn_\gamma}{dE_{\rm IR}}(E_{\rm IR},0)
 =
 \frac{N_\gamma\Gamma_X n_{X,c}(z_*)}
 {E_{\rm IR}H(z_*)}
 \Theta(E_*-E_{\rm IR}),
 \label{eq:decay-spectrum}
\end{equation}
where $z_*$ is the decay redshift selected by
$1+z_*=E_*/E_{\rm IR}$, and $n_{X,c}$ is the comoving parent abundance,
including any depletion.  The corresponding intensity is
$\nu I_\nu=E_{\rm IR}^2[dn_\gamma/dE_{\rm IR}](E_{\rm IR},0)/(4\pi)$
in natural units.
Eq.~\eqref{eq:decay-spectrum} displays the transparent-propagation limit
and is useful for understanding the spectral shape.  In obtaining the limits
below, we retain the redshift dependence and depletion of the parent
population, add the resulting photon density to the target field in
Eq.~\eqref{eq:optical-depth}, and constrain the extra opacity in $\tau^{\rm NP}_\gamma$.  We quote
separate 95\% exclusions based on the broad-band EBL reconstruction of
Gr\'eaux et al.~\cite{Greaux:2024} and on the LHAASO spectrum of
GRB~221009A~\cite{LHAASO:2023}.

\subsection{Active neutrino dipole moments}

\begin{figure}[ht]
 \centering
 \includegraphics[width=\columnwidth]{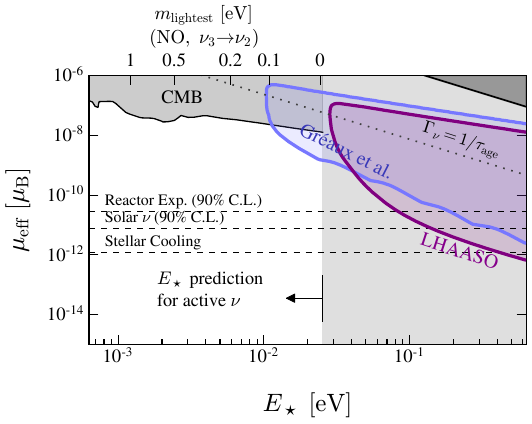}
 \caption{95\% confidence exclusions for the effective neutrino transition
 moment versus rest-frame energy $E_*$. The top axis gives the lightest mass
 for normal ordering and $\nu_3\to\nu_2$. Purple: LHAASO~\cite{LHAASO:2023};
 blue: Gr\'eaux et al.~\cite{Greaux:2024}. Known active neutrinos require
 $E_*\leq0.025\,\mathrm{eV}$; larger energies represent a generic spectrum.
 Dashed lines: GEMMA~\cite{Beda:2009kx,Beda:2012zz}, solar neutrinos~\cite{Borexino:2017fbd,Giunti:2023yha},
 and stellar cooling~\cite{Raffelt:1999gv,Viaux:2013hca,Viaux:2013lha,Capozzi:2020cbu}. Gray: CMB~\cite{Mirizzi:2007jd,Aalberts:2018obr}.}
 \label{fig:neutrino-constraints}
\end{figure}

For the radiative transition $\nu_i\to\nu_j\gamma$ {through $\mathcal{L}\supset \frac{1}{2}\bar{\nu}^j\sigma_{\mu\nu}(\mu_{ij}+\epsilon_{ij}\gamma_5)\nu^i F^{\mu\nu} \,+\,\text{h.c.}$}, where $i,j$ are the labels for the mass eigenstates, we define
$\mu_{\rm eff}^2=|\mu_{ij}|^2+|\epsilon_{ij}|^2$. $\mu_{ij}$ and
$\epsilon_{ij}$ are the magnetic and electric transition moments in the mass
basis, {and note that this operator flips chirality between the left- and right-handed states.}
The emitted photon energy and decay width are
\begin{equation}
 E_* = \frac{m_i^2-m_j^2}{2m_i},\qquad
 \Gamma_{i\to j}=\frac{\mu_{\rm eff}^2}{8\pi}
 \left(\frac{m_i^2-m_j^2}{m_i}\right)^3.
 \label{eq:neutrino-radiative-decay}
\end{equation}
For the standard relic abundance, Fig.~\ref{fig:neutrino-constraints} shows
our TeV-transparency limit on $\mu_{\rm eff}$.  For transitions among the three active
mass eigenstates, the atmospheric splitting implies
$E_*\leq\sqrt{|\Delta m_{31}^2|}/2\simeq0.025\,\mathrm{eV}$; the curves at
larger $E_*$ therefore apply only to a more general neutrino spectrum.
For the normal-ordering $\nu_3\to\nu_2$ transition shown here, the
conservative Planck+BAO bound $\sum_i m_i<0.12\,\mathrm{eV}$ in flat
$\Lambda$CDM corresponds approximately to
$m_{\rm lightest}\lesssim0.03\,\mathrm{eV}$~\cite{Planck:2018vyg}.
In this mass range, the lower edge of our conditional residual-EBL
exclusion band lies below the COBE/FIRAS spectral-distortion
bound~\cite{Mirizzi:2007jd,Aalberts:2018obr}.
While reactor and solar-neutrino scattering and stellar cooling provide stronger
constraints~\cite{Beda:2009kx,Beda:2012zz,Borexino:2017fbd,Giunti:2023yha,Raffelt:1999gv,Viaux:2013hca,Viaux:2013lha,Capozzi:2020cbu} (see also \cite{Giunti:2024gec} for a comprehensive review and \cite{Bernal:2021ylz} for a proposed cosmological probe using line-intensity mapping),
we note that these measurements probe different effective combinations of dipole
moments, whereas radiative decay selects a particular mass-eigenstate
transition \cite{Ternes:2025lqh}.  Therefore, these probes are complementary.

\subsection{Sterile-neutrino decay}

For $\nu_s\to\nu\gamma$, the rest-frame photon energy is
\begin{equation}
 E_*=\frac{m_{\nu_s}^2-m_\nu^2}{2m_{\nu_s}}
 =\Delta m\left(1-\frac{\Delta m}{2m_{\nu_s}}\right),
 \quad \Delta m=m_{\nu_s}-m_\nu .
 \label{eq:sterile-radiative-decay}
\end{equation}
We present the result directly in terms of the radiative width
$\Gamma_{\rm sterile}$ for initial sterile-neutrino fractions
$f_{\rm sterile}\equiv\rho_{\rm sterile}/\rho_{\rm DM}=10^{-2},10^{-4},10^{-6}$ at recombination.
A notable feature of this channel is its sensitivity to a highly subdominant
population: even when sterile neutrinos constitute far below one percent of
the dark matter, their decay photons accumulate over cosmological time and
can produce measurable TeV attenuation.  The usual structure-formation
limits are simultaneously relaxed in this regime.  The small-scale power
suppression probed by the Lyman-$\alpha$ forest decreases with the warm
fraction \cite{Boyarsky:2009ix}, while the Tremaine--Gunn phase-space mass
floor scales approximately as $f_{\rm sterile}^{1/4}$ when the sterile component supplies
only a fraction of the halo density \cite{Tremaine:1979we} {(see also \cite{Cline:2026tkp} for a recent discussion of light fermionic dark matter).}  Thus, at or below
the percent level, neither bound should be imported from an all-dark-matter
sterile-neutrino scenario without rescaling.
\begin{figure}[ht]
 \centering
 \includegraphics[width=\columnwidth]{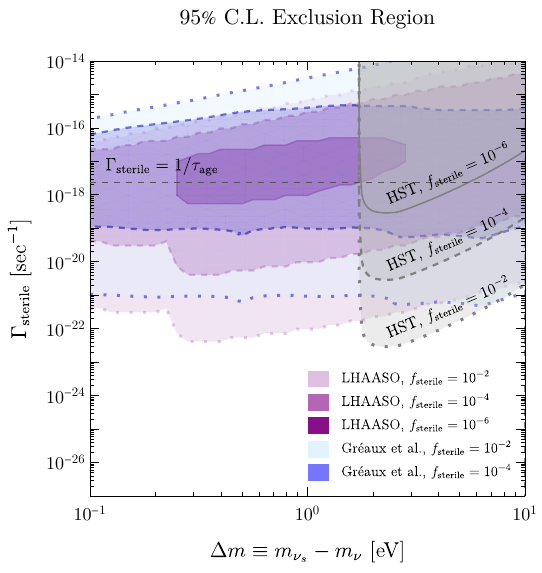}
 \caption{95\% confidence exclusions for the sterile-neutrino radiative width
 versus mass splitting. Purple: LHAASO~\cite{LHAASO:2023};
 blue: Gr\'eaux et al.~\cite{Greaux:2024}. Only the closed shaded regions
 are excluded. Fractions in the legend are defined at recombination;
 {$m_\nu=0.05\,\mathrm{eV}$ is used as a benchmark
 },
 and the radiative branching ratio is unity.
 {The gray contours show HST constraints from cosmic optical background anisotropies \cite{Nakayama:2022jza,Carenza:2023qxh}.}
}
 \label{fig:sterile-constraints}
\end{figure}

Cold parents at recombination are an input to the decay calculation. We remain agnostic to its
production mechanism.  Pauli
exclusion nevertheless sets a minimum momentum spread at fixed abundance.
The check in Table~\ref{tab:sterile-phase-space} of the Supplemental Material shows
that a distribution with $p_F/m_{\nu_s}\leq0.1$ at recombination is possible only
for $\Delta m\gtrsim1.35,\,0.393,\,0.090\,\mathrm{eV}$ for the three
fractions above, respectively.  Smaller splittings require assessing
the parent momenta during the emission history.

Unlike a simple upper limit, the exclusions in
Fig.~\ref{fig:sterile-constraints} form closed regions.  At small widths too
few photons are produced, whereas at large widths the parent population
decays before the epochs to which the EBL measurement is most sensitive.  The
bound therefore depends on both the initial abundance and its subsequent
depletion.

\subsection{Axion decay}

\begin{figure}[ht]
 \centering
 \includegraphics[width=\columnwidth]{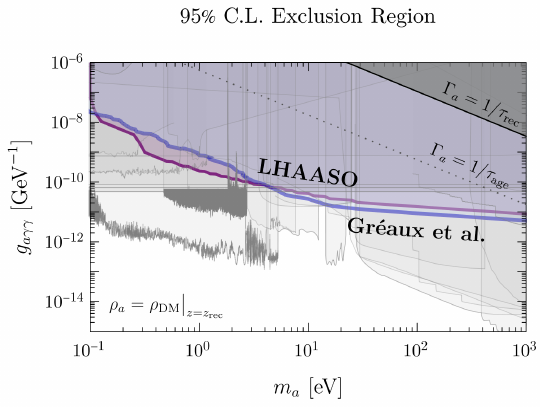}
 \caption{95\% confidence exclusions for the axion--photon coupling
 (regions above the boundaries). Purple: LHAASO~\cite{LHAASO:2023};
 blue: Gr\'eaux et al.~\cite{Greaux:2024}. Gray: existing constraints~\cite{Bernal:2023baf,Janish:2023jwst,Pinetti:2025jwst,Saha:2025jwst,Todarello:2023muse,Todarello:2024uv,Yin:2024winered,Grin:2006vimos,Wang:2023desi,Nakayama:2022jza,Carenza:2023qxh,Cadamuro:2012Cosmology,Arias:2012WISPy,Bolliet:2021PhotonInjection,Liu:2023EnergyInjection,PorrasBedmar:2024CosmicBackground,Langhoff:2022AxionBackground,Ayala:2014pea,Dolan:2022kul,Gondolo:2008dd,Nguyen:2023czp,Caputo:2021rux,Caputo:2022mah,Wadekar:2022LeoT,CAST:2007jps,CAST:2008ixs,CAST:2011rjr,CAST:2013bqn,CAST:2015qbl,Moriyama:1998kd,Inoue:2002qy,SAPPHIRES:2022bqg,SAPPHIRES:2024xmw,OHare:2020AxionLimits}.
 The gray comparisons retain their original abundance assumptions and
 confidence conventions.
 Axions initially equal the dark-matter abundance at recombination {in this work}.
  }
 \label{fig:axion-constraints}
\end{figure}

For an axion coupled to electromagnetism through
$\mathcal{L}\supset-g_{a\gamma\gamma}aF_{\mu\nu}\widetilde
F^{\mu\nu}/4$, each decay produces two photons ($N_\gamma=2$), with
\begin{equation}
 E_*=\frac{m_a}{2},\qquad
 \Gamma_{a\to\gamma\gamma}=\frac{g_{a\gamma\gamma}^2m_a^3}{64\pi}.
 \label{eq:axion-radiative-decay}
\end{equation}
We assume axions constitute all dark matter at recombination. The predicted
photon density scales with the initial axion abundance and with
$g_{a\gamma\gamma}^2$.
In Fig.~\ref{fig:axion-constraints}, the solid black and dotted gray lines mark decay rates
equal to the inverse recombination time and the inverse age of the Universe,
respectively. 
{
Following the $\gamma$-ray attenuation approach of \cite{Bernal:2023baf} but with updated data that are consistent across different EBL measurements, we also show a competitive constraint around $m_a\simeq 10\,{\rm eV}$.
}

\FloatBarrier
\section{Conclusion}
In this Letter, we demonstrate that TeV transparency, as an independent state-of-the-art measurement EBL in the $10^{-2} - 10$ eV energy range, can be a powerful probe of new physics beyond the standard model. As it provides {one of the most stringent measurements} of EBL to-date, any new physics that injects photons that fall into this band today can be constrained. {We derive strong constraints on the active-neutrino electromagnetic transition moment, the sterile-neutrino radiative decay rate, and the axion-photon coupling.} 
{We note that while the intrinsic spectra  of TeV gamma-ray sources remain under study, hiding a new-physics EBL contribution would require source-specific exponential hardening for various TeV gamma-ray sources, which is hard.} On the other hand, a better measurement of EBL and an improved modeling of the TeV gamma ray sources would most definitely improve the constraint we derive here. 

\textbf{Acknowledgement}~~~ We thank Yifan Chen, Chao-Qiang Geng, Tao Liu, Andrew Long, Mehrdad Mirbabayi, Giovanni Villadoro, Hyung Do Kim, Hang Bae Kim, and Yue-Liang Wu for useful communications. 
HS was supported by the Deutsche Forschungsgemeinschaft under
Germany Excellence Strategy — EXC 2121 “Quantum
Universe” — 390833306. 
HS appreciates the financial support and warm hospitality of the CERN Theory Group during the visit.
CS is partly supported by National Natural Science Foundation of China (NSFC) under Grant No. 12547104.
SY is supported by Basic Science Research Program through the National Research Foundation of Korea(NRF) funded by the Ministry of Education(No. RS-2026-25571203) and by IBS under the project code, IBS-R018-D1.
HS and CS thank the organizers of the workshop ``Axion-Photon Conversion: Terrestrial and Cosmic Searches", during which stimulating discussions and interesting program helped improved this work. 
HS and SY thank APCTP, Pohang, Korea for their hospitality during the Focus Program [APCTP-2026-F01] from which this work greatly benefited.

\FloatBarrier
\clearpage
\makeatletter
\let\supplement@sectioncntformat\@sectioncntformat
\let\supplement@hangfrom@section\@hangfrom@section
\appendix
\let\@sectioncntformat\supplement@sectioncntformat
\let\@hangfrom@section\supplement@hangfrom@section
\renewcommand{\thesection}{\Roman{section}}
\def\theequation@prefix{\Alph{section}}
\makeatother
\setcounter{secnumdepth}{2}
\begin{widetext}
\addtocontents{toc}{\protect\supplementaltoctrue}
\begin{center}
\textbf{Supplemental Material}
\end{center}
\supplementalcontents
\end{widetext}

\section{Optical Depth and EBL Number Density}

We use the energy convention of the main text: $E_{\rm UV}$ is the
gamma-ray energy observed \textit{today}, and $E_{\rm IR}$ is the local EBL photon
energy \textit{at redshift $z$}.  The density $dn_\gamma/dE_{\rm IR}$ is per unit
proper volume and local energy.  At $z=0$, $E_{\rm IR}$ is also the energy
measured today.

For an isotropic background, a gamma ray crossing a proper distance
$d\ell$ accumulates optical depth
\begin{align}
 d\tau_\gamma={}&d\ell\int_{-1}^{1}d\mu\,\frac{1-\mu}{2}
 \nonumber\\[-2pt]
 &\times\int_{E_{\rm IR}^{\rm th}}^\infty dE_{\rm IR}\,
 \frac{dn_\gamma}{dE_{\rm IR}}(E_{\rm IR},z)
 \sigma_{\gamma\gamma}(s),
 \label{eq:appendix-dtau}
\end{align}
where $\mu=\cos\theta$, $s=2(1+z)E_{\rm UV}E_{\rm IR}(1-\mu)$, and
$E_{\rm IR}^{\rm th}=2m_e^2/[(1+z)E_{\rm UV}(1-\mu)]$, as in
Eq.~\eqref{eq:optical-depth}.  The factor $1-\mu$ comes from the
relativistic collision flux: restoring $c$, the collision rate contains
$c(1-\mu)$, whereas the gamma ray travels $d\ell=c\,dt$.  The factor $1/2$
comes from averaging directions in an isotropic background after integrating
over the azimuthal angle.  The survival probability is $e^{-\tau_\gamma}$.

Along the photon path, a redshift interval corresponds to proper distance
\begin{equation}
 d\ell=|dt|=\frac{dz}{(1+z)H(z)}.
 \label{eq:appendix-line-element}
\end{equation}
Substituting this path length into Eq.~\eqref{eq:appendix-dtau} and
integrating from $z=0$ to the source redshift $z_s$ gives
Eq.~\eqref{eq:optical-depth}.

The same photon density determines the spectral intensity $\nu I_\nu$
used in the EBL comparison:
\begin{equation}
 [\nu I_\nu](E_{\rm IR},z)
 =\frac{E_{\rm IR}^2}{4\pi}
 \frac{dn_\gamma}{dE_{\rm IR}}(E_{\rm IR},z).
 \label{eq:appendix-number-density}
\end{equation}
The two powers of $E_{\rm IR}$ convert photon number to energy and a
linear energy interval to a logarithmic interval.  Isotropy supplies the
factor $1/(4\pi)$.  Restoring ordinary units multiplies the right-hand side
by $c$.  The residual EBL data constrain this intensity at $z=0$.

For the decay contribution, we use $E_*$, $N_\gamma$, $\Gamma_X$, and the
comoving parent density $n_{X,c}$ already defined in the main text.  A photon
emitted at redshift $z_*$ with energy $E_*$ has local energy $E_{\rm IR}$
at redshift $z$.  At fixed $z$, its redshift relation gives
\begin{align}
 1+z_*={}&(1+z)\frac{E_*}{E_{\rm IR}},
 \nonumber\\[-2pt]
 \left|\frac{dt(z_*)}{dE_{\rm IR}}\right|
 ={}&\frac{1}{E_{\rm IR}H(z_*)}.
 \label{eq:appendix-decay-redshift}
\end{align}
Here $t(z_*)$ is cosmic time at emission.  The number of photons produced
per comoving volume in an emission interval is
$N_\gamma\Gamma_X n_{X,c}(z_*)\,dt(z_*)$.  Converting emission time to local
photon energy with Eq.~\eqref{eq:appendix-decay-redshift}, and comoving to
proper volume with $(1+z)^3$, gives
\begin{equation}
 \frac{dn_\gamma}{dE_{\rm IR}}(E_{\rm IR},z)
 =
 \frac{(1+z)^3N_\gamma\Gamma_X n_{X,c}(z_*)}
 {E_{\rm IR}H(z_*)}
 \Theta(E_*-E_{\rm IR}).
 \label{eq:appendix-decay-density}
\end{equation}
The step function restricts emission to $z_*\geq z$.  The parent density
includes depletion and is set to zero outside the modeled parent history.
At $z=0$, this reduces to Eq.~\eqref{eq:decay-spectrum}; see also
Ref.~\cite{Bernal:2023baf}.  This is the photon density used in the
decay-opacity calculation.

\section{Methodology}
\label{sec:methodology}

We compare the decay signal separately with the present-day residual EBL
and the LHAASO gamma-ray spectra.  The predictions follow from
Eqs.~\eqref{eq:appendix-number-density} and
\eqref{eq:appendix-decay-density}, using the parent abundances specified
for each particle scenario.

\subsection{The Greaux data set}

We use eight residual intensity points at $z=0$ from the EBL reconstruction
of Gr\'eaux et al.~\cite{Greaux:2024}, in units of
$\mathrm{nW\,m^{-2}\,sr^{-1}}$.  The residual is the gamma-ray-inferred
total EBL minus the integrated galaxy light (IGL) at the corresponding
wavelengths.  The IGL estimates of Driver et al.~\cite{Driver:2016} and
Koushan et al.~\cite{Koushan:2021} integrate measured galaxy fluxes and
number counts from surveys including GAMA, COSMOS, DEVILS, and Hubble
deep fields, with infrared observations from Spitzer and Herschel.
They are observational estimates with fitted count distributions and
faint-source extrapolations.  We use the digitized, already-subtracted
residuals and their asymmetric errors.

For each parameter point, we average the predicted decay intensity
$[\nu I_\nu]_{\rm NP}(E_{\rm IR},0)$ over each energy bin: the intensity
is integrated over energy and divided by the bin width.  The bins are
centered on the digitized energy points and share a logarithmic width
equal to their mean logarithmic spacing.  We retain the signed residuals,
including negative central values, when comparing them with the
non-negative decay signal.  This comparison uses the reconstructed EBL
points without refitting the underlying STeVECat spectra.

\subsection{LHAASO gamma ray data}

For GRB~221009A at $z_s=0.151$, we use the digitized high-energy
subset of the LHAASO spectra~\cite{LHAASO:2023}: six flux points from
230--300~s and seven from 300--900~s after the trigger, spanning
approximately 4--12~TeV.

The conventional EBL represents stellar emission and dust reprocessing
accumulated over cosmic history.  We adopt the Saldana-Lopez et
al.~\cite{SaldanaLopez:2021} template, constructed from HST/CANDELS galaxy
photometry using stellar and dust emission templates and completeness
corrections, with the LHAASO-fitted rescaling.  Its multiplicative factors
are $1.30$, $1.20$, and $0.40$ for wavelengths below $8\,\mu\mathrm m$,
between $8$ and $28\,\mu\mathrm m$, and above $28\,\mu\mathrm m$,
respectively~\cite{LHAASO:2023}.  We interpolate the central
present-day curve logarithmically in energy and intensity.  At redshift
$z$, we evaluate that curve at $E_{\rm IR}/(1+z)$ and multiply
the intensity by $(1+z)^4$.  The intensity is set to zero when
$E_{\rm IR}/(1+z)$ lies outside the tabulated energy range.
This approximates the background evolution by photon
redshifting and dilution, without additional galaxy emission.

The predicted flux in each observing interval is
\begin{equation}
 \Phi_{\rm pred}(E_{\rm UV})
 =\Phi_{\rm int}(E_{\rm UV})
 e^{-\tau_\gamma^{\rm gal}-\tau_\gamma^{\rm NP}},
 \label{eq:method-lhaaso-flux}
\end{equation}
where both optical depths are evaluated at $(E_{\rm UV},z_s)$ using
Eq.~\eqref{eq:optical-depth}.  We evaluate the digitized intrinsic curves
by linear interpolation at the observed energies.  These curves and
the conventional EBL remain fixed as the decay parameters vary.

\subsection{Comparison statistic and construction of contours}

For either comparison, we calculate
\begin{equation}
 \chi^2=\sum_i\frac{(M_i-y_i)^2}{\sigma_i^2},
 \label{eq:method-statistic}
\end{equation}
where $y_i$ is the measured intensity or flux and $M_i$ its prediction.
We use the upper error when $M_i\geq y_i$ and the lower error otherwise,
treating the supplied errors as independent.  For LHAASO, the statistic
is summed over both observing intervals.  The two data comparisons
produce separate contours and are not combined into a joint statistic.

At each scanned parameter point, we form
\begin{equation}
 \Delta\chi^2_0=\chi^2-\chi^2_{\rm no\,decay}.
 \label{eq:method-delta-statistic}
\end{equation}
The reference is zero added intensity for the residual EBL and zero added
opacity for LHAASO.  We display the numerical threshold
$\Delta\chi^2_0=3.84146$.  

The particle-specific scans use logarithmic parameter grids over the
ranges shown and retain the imposed lifetime restriction
$\Gamma_X[t(1100)-t(10^7)]\leq1$, where $t(z)$ is cosmic time.
Points outside this restriction are masked rather than excluded by the
data.  The generic scan described below omits this lifetime restriction.

\section{Generic constraints on monochromatic photon injections}
\label{sec:generic-decay}

\begin{figure*}[t]
 \centering
 \includegraphics[width=\textwidth]{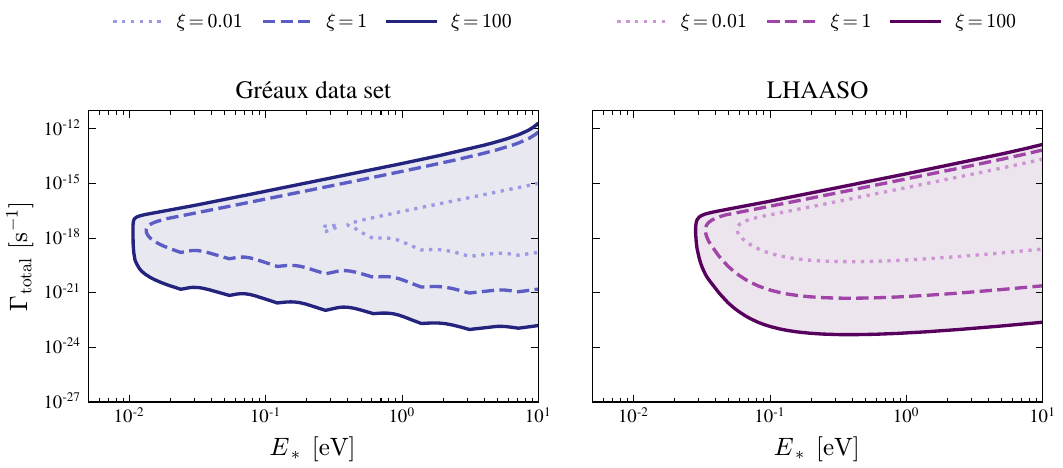}
 \caption{Generic decay constraints for $\xi=10^{-2},1,10^2$ at
 $z_{\rm ref}=1100$ [Eq.~\eqref{eq:generic-xi}].
 Blue (left): residual EBL; purple (right): LHAASO.
 Dotted, dashed, and solid curves denote increasing $\xi$,
 with darker shades for larger values. Curves mark
 $\Delta\chi^2_0=3.84146$; shading shows the region above threshold
 for $\xi=10^2$. Background EBL and source spectra are fixed,
 with no lifetime cut. }
 \label{fig:generic-decay}
\end{figure*}

The decay calculation can be expressed in terms of three parameters: the
photon energy at emission $E_*$, the total parent decay rate $\Gamma$, and
the dimensionless photon-weighted abundance
\begin{equation}
 \xi\equiv
 \frac{N_\gamma B_\gamma n_{X,c}(z_{\rm ref})}
 {1\,\mathrm{cm}^{-3}},\qquad z_{\rm ref}=1100.
 \label{eq:generic-xi}
\end{equation}
Here $n_{X,c}$ is the comoving parent number density, $N_\gamma$ is the
number of photons at energy $E_*$ per radiative decay, and
$B_\gamma=\Gamma_{\rm rad}/\Gamma$ is the radiative branching fraction.
Thus $\xi$ measures the available photon yield per comoving volume at
recombination; it is neither a dark-matter fraction nor the surviving
parent abundance today.  Photon multiplicity, branching fraction, and
reference abundance are degenerate at fixed $\Gamma$.

For homogeneous, nonrelativistic parents with a constant decay rate,
freely propagating photons, and the fixed background cosmology used here,
Eq.~\eqref{eq:appendix-decay-density} becomes
\begin{align}
 \frac{dn_\gamma}{dE_{\rm IR}}(E_{\rm IR},z)
 ={}&\frac{(1+z)^3(1\,\mathrm{cm}^{-3})\xi\Gamma}
 {E_{\rm IR}H(z_*)}
 e^{-\Gamma[t(z_*)-t(z_{\rm ref})]}
 \nonumber\\[-2pt]
 &\times\Theta(z_*-z)\Theta(z_{\rm ref}-z_*),
 \label{eq:generic-spectrum}
\end{align}
where $1+z_*=(1+z)E_*/E_{\rm IR}$.  The decay contributions to both
observables therefore obey
\begin{equation}
 \bigl([\nu I_\nu]_{\rm NP},\tau_\gamma^{\rm NP}\bigr)
 =\xi\,\bigl([\nu I_\nu]_{\rm NP},\tau_\gamma^{\rm NP}\bigr)_{\xi=1}
 \quad\text{at fixed }(E_*,\Gamma).
 \label{eq:generic-scaling}
\end{equation}
One set of response tables in $(E_*,\Gamma)$ consequently suffices for all
abundances and particle models satisfying these assumptions.  The total
rate cannot generally be absorbed into $\xi$, since it also sets the
exponential depletion and hence the spectral shape.  Only when
$\Gamma[t(z_*)-t(z_{\rm ref})]\ll1$ throughout the relevant emission
history do the predictions depend on $E_*$ and the product $\xi\Gamma$
alone, even if $B_\gamma=1$.

Figure~\ref{fig:generic-decay} shows three fixed-$\xi$ slices of this
common constraint.  For each data set we evaluate
\begin{equation}
 \Delta\chi^2_0(E_*,\Gamma,\xi)
 =\chi^2(E_*,\Gamma,\xi)-\chi^2(\xi=0).
 \label{eq:generic-statistic}
\end{equation}
For the eight present-day residual EBL points, we average the prediction
over top-hat energy windows centered on each point.  Their common
logarithmic width is the mean logarithmic point spacing, and we use the
asymmetric errors.  The LHAASO comparison uses the 13 high-energy
flux points, with the conventional EBL opacity and intrinsic source
spectra held fixed.  Each normalization is applied to the intensity and
additional opacity before computing the statistic; neither $\chi^2$ nor
the attenuated flux scales linearly with $\xi$.
We display the nominal threshold $\Delta\chi^2_0=3.84146$, without a
nuisance-parameter refit or a confidence-level calibration.  

Where depletion is negligible, the low-rate boundaries move downward
in inverse proportion to $\xi$.  At larger rates, depletion removes
parents before the later emission needed to populate the observed bands,
producing an upper boundary.  The adopted top-hat EBL bins and their unequal
residuals and errors give the lower edges their \mbox{scalloped} shape and can
produce disconnected islands, as for \mbox{$\xi=10^{-2}$} in the left panel.

We specify the reference abundance independently of the parent lifetime.
The generic scan imposes no restriction
$\Gamma^{-1}>t(z_{\rm ref})$ and does not constrain the mechanism that
produces the reference population.  A population specified at an earlier
epoch $z_i$ is mapped using
$n_{X,c}(z_{\rm ref})=n_{X,c}(z_i)
e^{-\Gamma[t(z_{\rm ref})-t(z_i)]}$.
For the plotted range $E_*\leq10\,\mathrm{eV}$, the supplied EBL bin edges
and gamma-ray energies restrict contributing emission to $z_*<944$ and
$z_*<603$, respectively.  Photons emitted before recombination therefore
do not contribute in these bands, so this conversion also reproduces
the earlier-normalized active-neutrino calculation within its cold-parent
approximation.

For example, axions map through $E_*=m_a/2$ and
$\xi=2f_a\rho_{{\rm DM},0}/[m_a(1\,\mathrm{cm}^{-3})]$, whereas sterile
neutrinos with unit radiative branching map through
$E_*=(m_{\nu_s}^2-m_\nu^2)/(2m_{\nu_s})$ and
$\xi=f_{\rm sterile}\rho_{{\rm DM},0}/[m_{\nu_s}(1\,\mathrm{cm}^{-3})]$; the fractions here
are specified at recombination.  One active-neutrino parent eigenstate
with unit radiative branching uses $E_*=\Delta m^2/(2m_i)$ and
$\xi=113e^{-\Gamma[t(z_{\rm ref})-t(z_i)]}$ with $z_i=2\times10^9$.
The model-specific plots thus follow by substituting the appropriate
normalization and converting coordinates, without repeating the
cosmological and pair-production integrals.  Any particle decaying
into a monochromatic photon line under these assumptions must satisfy
this constraint, independently of its spin or decay operator.
Relativistic parents, broad spectra, and photon reprocessing require
extensions.

\section{Assumptions and self-consistency check}
\label{sec:assumptions}

We examine the consistency of the assumed parent populations and the
treatment of the spatial and spectral distributions of the EBL.

\subsection{Fermionic phase-space consistency of the sterile benchmarks}
\label{sec:sterile-phase-space}

The sterile-neutrino calculation specifies an abundance at
$z_{\rm ref}=1100$ and uses nonrelativistic parents, without deriving their
production or thermal history.  A necessary consistency check follows from
Pauli exclusion, even at zero temperature
\cite{Carena:2021degenerate}.  Let $F(\boldsymbol p)$ be the occupation per
internal state, with $0\leq F\leq1$, and take $g_s=2$ available states.
At fixed physical number density, the smallest momentum moments occur for
the filled Fermi sphere $F=\Theta(p_F-p)$:
\begin{equation}
 n_s=\frac{g_s}{(2\pi)^3}\int d^3p\,F(\boldsymbol p)
     =\frac{g_s p_F^3}{6\pi^2},\qquad
 p_{\rm rms,min}=\sqrt{\frac35}\,p_F.
 \label{eq:sterile-fermi-momentum}
\end{equation}
 A population with
$p_{\rm rms}\ll m_{\nu_s}$ requires $p_F/m_{\nu_s}\ll1$; a production spectrum with
lower occupation can be warmer than this minimum.

Bar et al.~\cite{Bar:2021fornax} used this minimum-dispersion
argument to show that nonthermal production alone cannot evade
cosmological free-streaming bounds, revisiting the proposal of Randall
et al.~\cite{Randall:2016fermi}.  Their structure-formation bound assumes
the fermions constitute the dark matter and is not applied directly to
our subdominant, decaying populations.

The normalization actually supplied to the decay calculation is
\begin{equation}
 n_s(z_{\rm ref})=
 \frac{f_{\rm sterile}\rho_{{\rm DM},0}}{m_{\nu_s}}(1+z_{\rm ref})^3,
 \qquad m_{\nu_s}=\Delta m+0.05\,\mathrm{eV}.
 \label{eq:sterile-reference-density}
\end{equation}
Here $f_{\rm sterile}$ normalizes the rest-mass density; it equals the energy-density
fraction $\rho_s/\rho_{\rm DM}$ only in the nonrelativistic regime.
Using $\rho_{{\rm DM},0}=1.264\times10^3\,\mathrm{eV\,cm^{-3}}$, consistent
with $\Omega_{\rm DM}h^2=0.120$, gives
\begin{align}
 \eta_F\equiv\frac{p_F(z_{\rm ref})}{m_{\nu_s}}
 ={}&0.1566\left(\frac{f_{\rm sterile}}{10^{-2}}\right)^{1/3}
       \left(\frac{2}{g_s}\right)^{1/3}
       \left(\frac{1\,\mathrm{eV}}{m_{\nu_s}}\right)^{4/3}
       \nonumber\\[-2pt]
 &\times\frac{1+z_{\rm ref}}{1101}.
 \label{eq:sterile-fermi-ratio}
\end{align}
For a chosen ratio $\eta$, the condition $\eta_F\leq\eta$ therefore requires
\begin{align}
 m_{\nu_s}\geq{}&0.2489\,\mathrm{eV}
       \left(\frac{f_{\rm sterile}}{10^{-2}}\right)^{1/4}
       \left(\frac{2}{g_s}\right)^{1/4}
       \nonumber\\[-2pt]
 &\times\left(\frac{1+z_{\rm ref}}{1101}\right)^{3/4}
       \eta^{-3/4}.
 \label{eq:sterile-coldness-threshold}
\end{align}
The case $\eta=1$ marks an order-one momentum, not a cold population.
We also give $\eta=0.1$ as an illustrative criterion: in the filled-sphere
distribution every particle then has $v<0.1$, and
$p_{\rm rms}/m_{\nu_s}\leq0.0775$.  

\begin{table}[t]
\centering
\begin{tabular}{c c c}
\hline\hline
$f_{\rm sterile}$ & $\Delta m_{\min}$ ($\eta=1$)
      & $\Delta m_{\min}$ ($\eta=0.1$)\\
      & [eV] & [eV]\\
\hline
$10^{-2}$ & $0.1989$ & $1.3497$\\
$10^{-4}$ & $0.02871$ & $0.3926$\\
$10^{-6}$ & $0$ & $0.08997$\\
\hline\hline
\end{tabular}
\caption{Minimum splittings allowing the indicated Fermi-momentum ratio
at recombination, for $g_s=2$ and $m_\nu=0.05\,\mathrm{eV}$.  We use
$\Delta m_{\min}=\max(0,m_{\nu_s}^{\min}-m_\nu)$; the zero entry means that
$\eta_F<1$ for every positive splitting.  These are applicability checks of
the cold-parent assumption, not additional observational exclusions.}
\label{tab:sterile-phase-space}
\end{table}

For example, $\Delta m=0.1\,\mathrm{eV}$ corresponds to
$m_{\nu_s}=0.15\,\mathrm{eV}$ and gives
$\eta_F=1.96,\,0.423,\,0.0912$ for
$f_{\rm sterile}=10^{-2},10^{-4},10^{-6}$, respectively.  The first benchmark cannot
have a deeply nonrelativistic population at recombination even in the
minimum-momentum configuration; the smallest fraction allows the chosen
$\eta_F\leq0.1$ criterion.  The thresholds concern the parent mass, so the
daughter mass must be added to the splitting plotted in
Fig.~\ref{fig:sterile-constraints}.

Passing the $\eta=0.1$ check permits the illustrative cold distribution but
does not show that a particular production mechanism creates it.  Taking
only one available internal state raises the parent-mass thresholds by $2^{1/4}$; a maximum
occupation $F_{\max}<1$ raises them by a further $F_{\max}^{-1/4}$.
If parents free-stream after the reference epoch, their momenta redshift
as $p\propto1+z$.  Momentum-independent decay reduces the occupation but
does not cool the surviving particles.  Conversely, a population warm at
recombination can become nonrelativistic before the later emissions
relevant to the EBL.  Assessing such a case requires its momentum-dependent
decay spectrum and survival history.  However, as we see in the check, we only expect $\mathcal O(1)$ correction even at as light as $m_{\nu_s} \sim 0.1 \,\mathrm eV$, while at $m_{\nu_s} \gtrsim 1\, \mathrm{eV}$ the correction is negligible.

\subsection{Inhomogeneous EBL contributions}

The finite propagation time of decay photons makes their instantaneous
distribution inhomogeneous.  For example, temporarily restoring $c$, a
continuously emitting source with daughter-photon luminosity $\dot N_\gamma$
produces
\begin{equation}
 n_\gamma(r,t)
 =\frac{\dot N_\gamma(t-r/c)}{4\pi c r^2}\Theta(ct-r).
 \label{eq:inhomogeneous-source-density}
\end{equation}
Nevertheless, spatial averaging over sources of number density $n_g$ gives
\begin{equation}
 \overline n_\gamma(t)
 =n_g\int 4\pi r^2\,dr\,n_\gamma(r,t)
 =n_g\int_0^t dt'\,\dot N_\gamma(t').
 \label{eq:inhomogeneous-mean-density}
\end{equation}
Regions of enhanced density are therefore compensated by regions of reduced
density.  Since the pair-production optical depth is linear in the target
photon phase-space density, averaging over random lines of sight gives
\begin{equation}
 \langle\tau\rangle=\tau[\overline n_\gamma].
 \label{eq:inhomogeneous-mean-tau}
\end{equation}
Thus, inhomogeneity does not change the ensemble-mean optical depth.

It does, however, affect the flux-averaged attenuation.  Defining
\begin{equation}
 \langle F_{\rm obs}\rangle
 =F_{\rm int}\langle e^{-\tau}\rangle
 \equiv F_{\rm int}e^{-\tau_{\rm eff}},
\end{equation}
convexity of the function implies
\begin{equation}
 \tau_{\rm eff}=-\ln\langle e^{-\tau}\rangle
 \leq\langle\tau\rangle.
 \label{eq:inhomogeneous-jensen}
\end{equation}
For small fluctuations,
\begin{equation}
 \tau_{\rm eff}
 =\langle\tau\rangle-\frac{1}{2}\operatorname{Var}(\tau)+\cdots,
\end{equation}
so clumpiness weakens the effective attenuation inferred from an ensemble of
observed fluxes.

To estimate the largest possible effect, consider an idealized model in
which every galaxy is surrounded by a uniform photon sphere of radius
$R_\gamma$, with no decay photons outside.  We normalize the density inside
the spheres so that the spatial mean, and hence $\overline\tau$, remains
fixed.  For Poisson-distributed galaxies, the mean number of spheres crossed
by a path of length $D$ is
\begin{equation}
 \lambda=n_g\pi R_\gamma^2D.
\end{equation}
Averaging over the chord length through each sphere gives
\begin{equation}
 \tau_{\rm eff}
 =\lambda\left[1-
 \frac{1-(1+2x)e^{-2x}}{2x^2}\right],
 \qquad x=\frac{3\overline\tau}{4\lambda}.
 \label{eq:inhomogeneous-tau-eff}
\end{equation}
In the limit of rare, individually opaque spheres,
$\tau_{\rm eff}\simeq\lambda$: most sightlines are transparent, while a
small fraction are completely absorbed.

Table~\ref{tab:inhomogeneous-ebl} shows this deliberately extreme estimate
for a galaxy radius $R_{\rm gal}=10\,\mathrm{kpc}$, number density
$n_g=10^{-2}\,\mathrm{Mpc}^{-3}$, and gamma-ray path length
$D=600\,\mathrm{Mpc}$.  The corresponding mean galaxy separation is
$d_g=n_g^{-1/3}=4.6\,\mathrm{Mpc}\simeq460R_{\rm gal}$.

\begin{table}[t]
 \centering
 \scriptsize
 \begin{tabular}{c c c c}
  \hline\hline
  $R_\gamma/R_{\rm gal}$ & $\lambda$
  & $1-\tau_{\rm eff}/\overline\tau$
    ($\overline\tau=1$)
  & $1-\tau_{\rm eff}/\overline\tau$
    ($\overline\tau=10$) \\
  \hline
  $1$   & $1.9\times10^{-3}$ & $99.81\%$ & $99.981\%$ \\
  $10$  & $0.188$             & $81.74\%$ & $98.12\%$  \\
  $30$  & $1.70$              & $26.59\%$ & $83.47\%$  \\
  $100$ & $18.85$             & $2.92\%$  & $24.43\%$  \\
  \hline\hline
 \end{tabular}
 \caption{Reduction of the flux-averaged effective optical depth in the
 uniform photon-sphere model.  The spatially averaged optical depth
 $\langle\tau\rangle=\overline\tau$ is unchanged.}
 \label{tab:inhomogeneous-ebl}
\end{table}

This toy model greatly overestimates the physical inhomogeneity of a freely
propagating decay background.  A radius of $1\,\mathrm{Mpc}$ corresponds to
only $3.3\,\mathrm{Myr}$ of propagation, whereas decay photons relevant to
the cosmological EBL generally accumulate for hundreds of Myr or several
Gyr.  After $1\,\mathrm{Gyr}$ their propagation radius is approximately
$300\,\mathrm{Mpc}$; the benchmark path then samples contributions from
$\lambda\simeq1.7\times10^6$ source spheres.  Equation
\eqref{eq:inhomogeneous-tau-eff} gives a fractional reduction of only
$3\times10^{-7}$ for $\overline\tau=1$ and
$3\times10^{-6}$ for $\overline\tau=10$.  A substantial correction
therefore requires either a photon horizon below a few Mpc, as could occur
extremely close to the pair-production threshold, or a rare and transient
source population.

Finally, the target field near an individual galaxy is anisotropic.  Daughter
photons streaming outward from the gamma-ray host are approximately
co-propagating with an escaping TeV photon, suppressing both the
$(1-\cos\theta)$ collision factor and the pair-production rate.  Positive
source clustering can instead enhance the conditional optical depth near the
host, but only over the galaxy correlation length, which is short compared
with a cosmological gamma-ray path.  These endpoint and angular corrections
are therefore expected to be subleading.  The homogeneous EBL calculation
should consequently be an excellent approximation for the mean opacity in
the parameter region considered here.

\subsection{Dependence on the spectral shape}
The attenuation of a gamma ray depends on the spectral distribution of the
EBL, rather than only on its bolometric intensity, because pair production
selects a finite range of target-photon wavelengths at each gamma-ray energy.
It is therefore important that the spectral basis used in the EBL
reconstruction can also represent the nonstandard component considered here.
Gr\'eaux et al.~\cite{Greaux:2024} parametrized the present-day EBL with eight
overlapping Gaussian basis functions in logarithmic wavelength,
\begin{equation}
 \nu I_\nu(\ell)=\sum_{i=1}^{8}a_i
 \exp\!\left[-\frac{(\ell-\ell_i)^2}{2\sigma^2}\right],
 \qquad \ell=\ln\!\left(\frac{\lambda}{\lambda_{\rm ref}}\right).
 \label{eq:ebl-gaussian-basis}
\end{equation}
The centers $\ell_i$ and common width $\sigma$ are fixed, while the eight
amplitudes $a_i$ are fitted independently.  For the illustrations below we
use $\lambda_i=0.3\,\mu\mathrm{m}\,e^{(i-1)\Delta\ell}$, with
$\Delta\ell=2\sigma\sqrt{2\ln 2}$, $\sigma/\ln 10=0.15$, and
$\lambda_{\rm ref}=0.6\,\mu\mathrm{m}$.  The eight centers then span
$0.3$--$89.1\,\mu\mathrm{m}$; the fit covers the quoted response interval
$0.18$--$120\,\mu\mathrm{m}$.  This illustrative grid is offset from the
published analytic prescription, which anchors a Gaussian at
$0.6\,\mu\mathrm{m}$.  The overlap values below refer to the explicitly
specified grid used in both figures.  This construction allows
the EBL intensity to vary locally in wavelength, although its finite basis
width limits the resolution with which a narrow spectral feature can be
reconstructed.

We test this explicitly using two physical decay spectra generated from
Eq.~\eqref{eq:appendix-decay-density}, including cosmological redshifting
and parent depletion.  For neutrinos we take normal ordering,
$\nu_3\to\nu_2\gamma$, $E_*=0.02\,\mathrm{eV}$, and
$\mu_{\rm eff}=10^{-8}\mu_B$.  The corresponding masses are
$(m_1,m_2,m_3)=(0.03465,0.03571,0.06093)\,\mathrm{eV}$, and the radiative
width is $\Gamma_\nu=3.397\times10^{-20}\,\mathrm{s}^{-1}$.
We use the standard comoving parent abundance
$113\,\mathrm{cm}^{-3}$ at $z_i=2\times10^9$ and one photon per decay.
For axions we take $m_a=5\,\mathrm{eV}$ and
$g_{a\gamma\gamma}=10^{-10}\,\mathrm{GeV}^{-1}$, giving
$E_*=2.5\,\mathrm{eV}$ and
$\Gamma_a=9.445\times10^{-24}\,\mathrm{s}^{-1}$.
Axions initially constitute all dark matter at $z_{\rm ref}=1100$ and
produce two photons per decay.  Both spectra are evaluated at $z=0$
in the nonrelativistic-parent approximation.

\begin{figure*}[t]
 \centering
 \includegraphics[width=\textwidth]{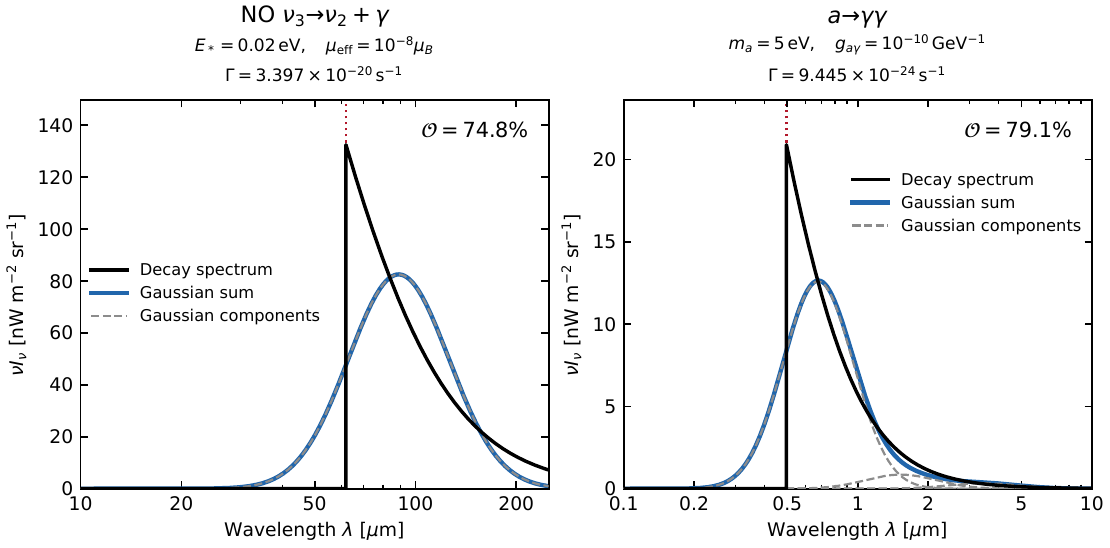}
 \caption{Non-negative eight-Gaussian fits (blue) to regenerated
 present-day decay spectra (black). Left: normal-ordering
 $\nu_3\to\nu_2\gamma$ with $E_*=0.02\,\mathrm{eV}$ and
 $\mu_{\rm eff}=10^{-8}\mu_B$, giving
 $\Gamma_\nu=3.397\times10^{-20}\,\mathrm{s}^{-1}$.
 Right: $m_a=5\,\mathrm{eV}$ and
 $g_{a\gamma\gamma}=10^{-10}\,\mathrm{GeV}^{-1}$, giving
 $\Gamma_a=9.445\times10^{-24}\,\mathrm{s}^{-1}$.
 Dashed gray curves show the nonzero fitted Gaussian components;
 the neutrino component coincides with its fitted sum. Dotted red lines
 mark the endpoints. The overlaps in Eq.~\eqref{eq:spectral-overlap}
 use the full $0.18$--$120\,\mu\mathrm{m}$ response interval, including
 wavelengths below each endpoint. Both spectra include redshifting
 and depletion, with abundances and reference epochs specified in the text.}
\label{fig:axion-spectral-basis}
\end{figure*}

For each spectrum $S(\lambda)$, shown in
Fig.~\ref{fig:axion-spectral-basis}, we use the quoted
$0.18$--$120\,\mu\mathrm{m}$ response range
and determine the amplitudes in Eq.~\eqref{eq:ebl-gaussian-basis} by
non-negative least squares, minimizing
$\int d\ln\lambda\,[S(\lambda)-B(\lambda)]^2$.  The target
is set to zero outside its kinematic support.  To quantify the agreement we
use the symmetric area-overlap coefficient
\begin{equation}
 \mathcal O
 =1-\frac{\int d\ln\lambda\,|S-B|}
 {\int d\ln\lambda\,(S+B)}
 =\frac{2\int d\ln\lambda\,\min(S,B)}
 {\int d\ln\lambda\,(S+B)},
 \label{eq:spectral-overlap}
\end{equation}
where $B(\lambda)$ is the fitted Gaussian sum.  Thus
$0\leq\mathcal O\leq1$, with unity corresponding to identical spectra,
including their normalization.

Over the full response interval we find
$\mathcal O_\nu=0.748$ and $\mathcal O_a=0.791$.
The respective endpoints are $\lambda_*=61.99\,\mu\mathrm{m}$ and
$0.496\,\mu\mathrm{m}$.  The neutrino fit has one nonzero Gaussian
component; the axion spectrum is represented by several components.
The broad basis follows the smooth continuum but leaks intensity to
wavelengths below the sharp endpoint, reducing the full-interval overlap.

The overlap also depends on the emitted photon energy $E_*$ because the
endpoint $\lambda_*=hc/E_*$ moves relative to the fixed Gaussian centers.
At fixed decay rate and reference epoch, changing $E_*$ translates the
normalized spectrum in logarithmic wavelength,
\begin{equation}
 \widehat S_{E_*,\Gamma}(\lambda)
 =\widehat S_{E_{*,\rm ref},\Gamma}\!\left(\lambda\frac{E_*}{E_{*,\rm ref}}\right),
 \label{eq:spectral-energy-scaling}
\end{equation}
where $\widehat S$ denotes intensity divided by its endpoint value, thus factorizing out the overall normalization.
We regenerate each spectrum and refit all eight non-negative amplitudes
at each $E_*$ over the same response interval.  This
parametrization applies to all three decay scenarios under the
nonrelativistic monochromatic-emission approximation: $E_*=m_a/2$ for
axions and $E_*=(m_i^2-m_f^2)/(2m_i)$ for active or sterile neutrinos.
Each curve keeps its decay rate and reference epoch fixed; an overall
intensity normalization cancels from $\mathcal O$ after the fit.

Changing the lifetime can additionally
change the shape through parent depletion.  In particular, writing
$J=\nu I_\nu$, constant-rate decays give the present-day intensity ratio
\begin{equation}
 \frac{J_{\rm dec}(E_{\rm IR},0)}{J_{\rm dec}(E_*,0)}
 =\frac{E_{\rm IR}}{E_*}\frac{H_0}{H(z_{\rm d})}
 e^{\Gamma[t_0-t(z_{\rm d})]},
 \label{eq:decay-normalized-shape}
\end{equation}
where $1+z_{\rm d}=E_*/E_{\rm IR}$, $t_0$ is the present cosmic age, and
the energies lie within the emission history's kinematic support.  Larger $\Gamma$ thus
enhances the redshifted tail relative to the endpoint once depletion is
important.  For $\Gamma t_0\ll1$, the exponential is approximately unity:
the shape is independent of $\Gamma$, which changes only the overall
normalization.  The overlap generally depends on both $E_*$ and $\Gamma$.
For the benchmarks above, $\Gamma_\nu t_0=0.0148$ and
$\Gamma_a t_0=4.11\times10^{-6}$, so both are close to the long-lived
regime.  The energy scan holds the quoted rates fixed; the couplings
are specified at the two benchmark energies and are not held fixed
along these curves.

\begin{figure}[t]
 \centering
 \includegraphics[width=\columnwidth]{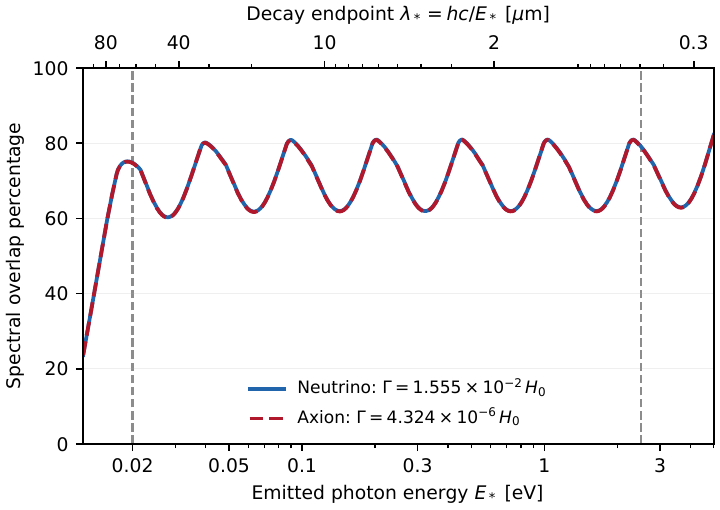}
 \caption{Present-day spectral overlap as a function of emitted photon
 energy $E_*$ at the two fixed decay rates used in
 Fig.~\ref{fig:axion-spectral-basis}, expressed in units of
 $H_0=67.4\,\mathrm{km}\,\mathrm{s}^{-1}\,\mathrm{Mpc}^{-1}$.
 At each $E_*$ the eight
 non-negative Gaussian amplitudes are refitted over
 $0.18$--$120\,\mu\mathrm{m}$, including wavelengths below the decay
 endpoint.  Gray dashed vertical lines mark the benchmark energies
 $E_*=0.02$ and $2.5\,\mathrm{eV}$; the upper axis gives the endpoint
 wavelength. The rates, reference epochs, Gaussian centers, and widths
 remain fixed along each curve.}
 \label{fig:decay-spectral-overlap-energy}
\end{figure}

Figure~\ref{fig:decay-spectral-overlap-energy} shows the dependence on
$E_*$ at the two benchmark rates.  The overlap varies non-monotonically
as the endpoint moves across the fixed basis.  Away from the response boundaries, the
modulation approximately repeats when $E_*$ changes by the Gaussian
spacing factor $e^{\Delta\ell}=2.26$.  Near the long-wavelength boundary,
the overlap falls further as the endpoint approaches
$120\,\mu\mathrm{m}$ ($E_*=0.0103\,\mathrm{eV}$).  Overall, we observe a good coverage of the chosen basis in Gr\'eaux et al.~\cite{Greaux:2024} to represent the monochromatic photon signals we discuss in this work. 

This overlap is evaluated at $z=0$: the fitted decay spectrum already
includes cosmological redshifting from its emission history.  A photon
emitted at redshift $z_{\rm d}$ with energy $E_*$ has local energy
$E_{\rm IR}=E_*(1+z)/(1+z_{\rm d})$ at a later epoch $z\leq z_{\rm d}$.
The Gaussian basis describes present-day wavelengths, rather than the
photon energy at emission.  Gr\'eaux et al.~\cite{Greaux:2024} evolve this
basis using $J\equiv\nu I_\nu$ and
\begin{equation}
 J_{\rm gal}(E_{\rm IR},z)
 =(1+z)^{4-f_{\rm evol}}
 J_{\rm gal}\!\left(\frac{E_{\rm IR}}{1+z},0\right),
 \label{eq:ebl-basis-evolution}
\end{equation}
with $1\leq f_{\rm evol}\leq2$.  Its centers in local wavelength are
therefore $\lambda_i/(1+z)$, with unchanged logarithmic widths.
For continuous decays, fresh photons instead keep the local upper energy
endpoint at $E_*$ at every epoch; the local spectrum follows
Eq.~\eqref{eq:appendix-decay-density}.  
The finite Gaussian basis captures a broad decay contribution with
energy-dependent accuracy.  For the constraints presented here we
consequently forward-model the decay photon density at each redshift in
the optical depth rather than identifying it with a single combination
of the reconstructed EBL amplitudes.

\subsection{Boosted parent particle and its impact}
\label{sec:boosted-parent}

A moving parent lives longer in cosmic time and emits photons over a
range of energies even from a two-body decay.  Both effects could change the EBL produced by its
decay.  For relic neutrinos, when the photons
probed by the EBL data were emitted matters: a large boost in the early Universe
need not imply a large correction to the EBL measured today.  Crudely speaking, the correction stays small as long as the neutrinos become non-relativistic before the at-rest decay time $1/\Gamma$.

We test
this for Fig.~\ref{fig:neutrino-constraints} by averaging over the standard
thermal relic momentum distribution, with isotropic photon emission in
the parent rest frame.  Thermal momentum effects in radiative neutrino
decay are also treated in Ref.~\cite{Aalberts:2018obr}.

Let $q$ denote the parent momentum redshifted to the present day.  The
initial number distribution is proportional to
$q^2dq/[\exp(q/T_{\nu0})+1]$, with
$T_{\nu0}=1.677\times10^{-4}\,\mathrm{eV}$ and
$\langle q\rangle=3.151T_{\nu0}=5.28\times10^{-4}\,\mathrm{eV}$.
At redshift $z$, its momentum and Lorentz factor are
\begin{equation}
 p(q,z)=q(1+z),\qquad
 \gamma(q,z)=\sqrt{1+\frac{q^2(1+z)^2}{m_i^2}}.
 \label{eq:boosted-parent-gamma}
\end{equation}
For the normal-ordering $\nu_3\to\nu_2$ transition used on the top axis of
Fig.~\ref{fig:neutrino-constraints}, $m_3\geq0.05011\,\mathrm{eV}$.
For an undepleted population at this minimum mass, the thermal average gives
$\langle\gamma\rangle=1.000072$ today, $1.000383$ at $z=1.3$, and
$11.67$ at recombination.  The parents therefore pass from relativistic
to very slow motion over their history.

Time dilation makes the decay rate per unit cosmic time
$\Gamma/\gamma$, where $\Gamma$ is the rest-frame width.  We follow the
survival of each momentum separately,
\begin{equation}
 S(q,z)=\exp\!\left[-\Gamma\int_z^{z_i}
 \frac{dz'}{(1+z')H(z')\gamma(q,z')}\right],
 \label{eq:boosted-parent-survival}
\end{equation}
starting at $z_i=2\times10^9$ with the same comoving abundance as the cold
calculation.
For the minimum parent mass, the accumulated proper time to today is,
on average, $33.6\,\mathrm{Myr}$ shorter than the cosmic elapsed time,
about $0.24\%$ of the age of the Universe.  Near the upper blue boundary,
where depletion is important, this leaves roughly $5\%$ more surviving
parents at fixed coupling.  The shift of the boundary itself is smaller,
as the decay signal there changes steeply with the coupling.

Motion also broadens the photon line.  A photon emitted at redshift $z_e$
at rest-frame angle $\theta_*$ to the parent's velocity is observed today
with energy
\begin{equation}
 E_{\gamma0}=\frac{E_*}{1+z_e}\gamma(q,z_e)
 \bigl(1+\beta(q,z_e)\cos\theta_*\bigr),
 \label{eq:boosted-photon-energy}
\end{equation}
where $\beta=p/\sqrt{m_i^2+p^2}$.  For isotropic emission, the mean
emitted energy is $\gamma E_*$, while individual photons can be shifted
up or down.  The early relativistic limit is especially instructive:
\begin{equation}
 E_{\gamma0}\longrightarrow
 \frac{E_*}{m_i}q(1+\cos\theta_*)\leq q.
 \label{eq:boosted-photon-early-limit}
\end{equation}
The inequality follows from $E_*/m_i\leq1/2$ for the two-body decay.
Cosmological redshift cancels the large early boost, leaving typical
photons well below the lowest EBL window, which starts at
$0.01059\,\mathrm{eV}$.  Reaching that window from the relativistic epoch
would require $q/T_{\nu0}\gtrsim63$, in the exponentially suppressed
tail of the relic distribution.

The active-neutrino signal in these EBL windows is consequently produced
late.  The maximum line energy for the displayed transition is
$E_*=0.02432\,\mathrm{eV}$, so the cold relation
$1+z_e=E_*/E_{\gamma0}$ selects $z_e\lesssim1.3$.  At these epochs the
small parent velocities mainly smooth the spectrum, rather than greatly
increase its photon energies.  We recompute the binned intensity and
additional gamma-ray opacity by integrating over the relic momenta and
emission angles, retaining the momentum-dependent survival in
Eq.~\eqref{eq:boosted-parent-survival}.  The data, initial abundance, and
statistic are held fixed.  Table~\ref{tab:boosted-neutrino-check} gives
representative changes of the blue boundaries.  A finer numerical
integration reproduces the shifts at $E_*=0.024\,\mathrm{eV}$.

\begin{table}[t]
\centering
\begin{tabular}{c c c}
\hline\hline
$E_*$ [eV] & Lower edge [\%] & Upper edge [\%]\\
\hline
$0.012$ & $-0.016$ & $+0.064$\\
$0.020$ & $-0.015$ & $+0.242$\\
$0.024$ & $+0.196$ & $+0.434$\\
\hline\hline
\end{tabular}
\caption{Changes in $\mu_{\rm eff}$ at the lower and upper edges of the
residual-EBL exclusion in Fig.~\ref{fig:neutrino-constraints}, including
thermal motion and time dilation for normal-ordering $\nu_3\to\nu_2$.
Entries are $100(\mu_{\rm boosted}/\mu_{\rm cold}-1)$; positive values
mean that the boundary moves to larger transition moments.}
\label{tab:boosted-neutrino-check}
\end{table}

For LHAASO, the maximum $\Delta\chi^2_0$ at the active-neutrino endpoint
remains $0.784$ after including boosts, below the adopted threshold
$3.84146$.  Thus this correction does not generate a purple exclusion
within the physical active-neutrino range.  The higher-$E_*$ extension of
Fig.~\ref{fig:neutrino-constraints} requires a separate choice of parent
mass and momentum distribution.  As a thermal example, we take
$m_i=2E_*$, corresponding to a massless daughter and the largest allowed
boost for fixed $E_*$ and relic momenta.  At $E_*=0.03\,\mathrm{eV}$,
the lower and upper LHAASO boundaries shift by $-0.77\%$ and $+0.33\%$;
at $0.1\,\mathrm{eV}$ the shifts are about $-0.001\%$ and $+0.10\%$.
These checks concern the stated thermal examples; a hotter or nonthermal
population requires specifying its momentum distribution.

\section{Decay-induced EBL versus direct axion absorption}

One might ask whether the same TeV photons could constrain axions directly,
without first converting the axions into an EBL photon population.  The
axion--photon coupling permits the $2\to2$ process
$\gamma_{\rm UV}a\to\gamma^*\to e^+e^-$.  For a nonrelativistic axion its
cross section is
\begin{equation}
 \sigma_{\gamma a}(s)=\frac{\alpha g_{a\gamma\gamma}^2}{12}
 \left(1-\frac{m_a^2}{s}\right)
 \sqrt{1-\frac{4m_e^2}{s}}
 \left(1+\frac{2m_e^2}{s}\right),
 \label{eq:direct-axion-absorption}
\end{equation}
where $s=m_a^2+2m_a(1+z)E_{\rm UV}$ at interaction redshift $z$, above
threshold.  The corresponding
interaction rate per TeV photon is
$R_{2\to2}=n_a\sigma_{\gamma a}$.

The process considered in this work is instead axion decay followed by
Breit--Wheeler scattering,
\begin{equation}
 a\to\gamma_{\rm IR}\gamma_{\rm IR},\qquad
 \gamma_{\rm UV}\gamma_{\rm IR}\to e^+e^-.
 \label{eq:decay-then-bw}
\end{equation}
It may be represented as the $2\to3$ reaction
$a+\gamma_{\rm UV}\to\gamma_{\rm IR}+e^+e^-$, but its dominant contribution
is not an ordinary local three-body process.  The internal photon can go on
shell: near $q^2=0$ the amplitude factorizes as
\begin{equation}
 \mathcal M_{2\to3}\simeq
 \mathcal M_{a\to\gamma\gamma}\frac{1}{q^2+i0}
 \mathcal M_{\gamma\gamma\to e^+e^-}.
 \label{eq:three-body-factorization}
\end{equation}
The on-shell photon can propagate for a cosmological time before interacting.
Consequently, the physical regulator of the pole is the finite propagation
time and geometry, and the appropriate description is production and
transport of real decay photons, as used in Eq.~\eqref{eq:decay-spectrum}.

This makes the two-step channel far more efficient than direct absorption.
In the long-lived regime, decay photons accumulate to a density
$n_{\gamma,\mathrm{dec}}\simeq2n_a\Gamma_a t$, where
$\Gamma_a=g_{a\gamma\gamma}^2m_a^3/(64\pi)$.  The ratio of the instantaneous
rates is therefore parametrically
\begin{equation}
 \frac{R_{\rm decay+BW}}{R_{2\to2}}
 \simeq 2\Gamma_a t\,
 \frac{\overline\sigma_{\gamma\gamma}}{\sigma_{\gamma a}}
 \sim \alpha\frac{m_a^3t}{m_e^2}.
 \label{eq:decay-direct-rate-ratio}
\end{equation}
Although only a small fraction of axions have decayed, each real daughter
subsequently interacts through the much larger QED Breit--Wheeler cross
section, with no additional axion-coupling suppression.  Both rates scale as
$n_ag_{a\gamma\gamma}^2$, so their ratio is independent of the axion abundance
and of $g_{a\gamma\gamma}$ while depletion is negligible.

For illustration, at $z=0$, $m_a=1\,\mathrm{eV}$, and
$E_{\rm UV}=10\,\mathrm{TeV}$,
$\Gamma_at_0=3.29\times10^{-8}g_{10}^2$, with
$g_{10}\equiv g_{a\gamma\gamma}/(10^{-10}\,\mathrm{GeV}^{-1})$, whereas the
spectrum-averaged Breit--Wheeler cross section is
$9.66\times10^{-26}\,\mathrm{cm}^2$, compared with
$\sigma_{\gamma a}\simeq2.37\times10^{-51}g_{10}^2\,\mathrm{cm}^2$.
The resulting local rate enhancement is about $2.7\times10^{18}$; including
redshift evolution along the line of sight to GRB~221009A gives an
optical-depth ratio of approximately $2.4\times10^{18}$.  Thus the
decay-generated EBL probed here overwhelmingly dominates the direct
$\gamma a$ channel.  The pole contribution and the decay-background opacity
are two descriptions of the same physical histories.

\section{Comparison of neutrino dipole probes}
\label{sec:neutrino-dipole-probes}

Different measurements constrain different combinations of neutrino dipole
moments.  We use the mass-basis matrices $\mu$ and $\epsilon$ for magnetic
and electric moments and define $\lambda=\mu-i\epsilon$.  These matrices
$\mu$ and $\epsilon$ are Hermitian; for Majorana neutrinos they are also
imaginary and antisymmetric, so only transition moments are allowed.
Dirac neutrinos can additionally have diagonal moments~\cite{Canas:2015dipole}.
The following scattering expressions assume ultrarelativistic neutrinos
and negligible spin conversion during propagation.

\emph{Reactor scattering.---}
Short-baseline reactor experiments such as GEMMA constrain the effective
moment of an initially coherent electron-antineutrino state, rather than
one mass-basis entry~\cite{Beda:2009kx,Grimus:1997dipole}:
\begin{equation}
 \mu_R^2=\sum_j\left|\sum_i U_{ei}
 (\mu_{ij}+i\epsilon_{ij})\right|^2,
 \label{eq:reactor-dipole-combination}
\end{equation}
where $U$ is the leptonic mixing matrix and the unobserved final mass
eigenstate $j$ is summed incoherently.  At finite oscillation distance,
$U_{ei}$ acquires the phase $\exp[-im_i^2L/(2E)]$.
Several dipole entries can interfere within each amplitude, making the
constraint sensitive to their relative phases.

\emph{Solar scattering.---}
Solar neutrinos reach the detector as an incoherent mixture of mass
eigenstates.  If $P_{ei}(E)$ is the probability of arriving in state $i$,
the effective scattering moment is
\begin{equation}
 \mu_S^2(E)=\sum_i P_{ei}(E)(\lambda^\dagger\lambda)_{ii}
 =\sum_{ij}P_{ei}(E)|\lambda_{ji}|^2.
 \label{eq:solar-dipole-combination}
\end{equation}
The measured recoil spectrum averages this expression over the solar
fluxes and detector response~\cite{Borexino:2017fbd}.
There is no interference between incoming mass eigenstates.  In
particular, for a single Majorana transition $i\leftrightarrow j$,
$\mu_S^2=(P_{ei}+P_{ej})\mu_{\rm eff}^2$, with $\mu_{\rm eff}$ defined
in Eq.~\eqref{eq:neutrino-radiative-decay}.  Thus solar data also bound
individual active transition moments; their limits cannot in general be
removed by choosing relative phases~\cite{Canas:2015dipole}.

\emph{CMB spectral distortions.---}
Radiative decay and inverse photon absorption distort the CMB spectrum.
For a specified transition, COBE/FIRAS constrains the same
$\mu_{\rm eff}^2=|\mu_{ij}|^2+|\epsilon_{ij}|^2$ that enters the decay
width in Eq.~\eqref{eq:neutrino-radiative-decay}, with sensitivity set by
the masses and relic populations.  The calculation of
Ref.~\cite{Aalberts:2018obr} includes thermal neutrino momenta, stimulated
emission, and Pauli blocking, assuming negligible depletion.

\emph{This work.---}
We constrain that same transition strength through its decay-generated
photon background: its present-day intensity is compared with the
Gr\'eaux residual EBL, and its additional pair-production optical depth is
compared separately with LHAASO spectra.  Our cold-parent calculation
uses the standard relic abundance and a decay-survival factor, with
thermal-motion checks in Sec.~\ref{sec:boosted-parent}.  For the physical
normal-ordering $\nu_3\to\nu_2$ range in
Fig.~\ref{fig:neutrino-constraints}, the residual-EBL comparison supplies
the exclusion band.  Complementarity with CMB distortions therefore
comes from the photon energies and redshifts probed, rather than a
different dipole combination.  Several contributing decay channels
would require consistent branching fractions and population evolution.
Although scattering and decay probe different effective combinations,
this distinction alone does not evade the stronger solar and stellar
limits on ordinary active-neutrino transitions.

\addtocontents{toc}{\protect\supplementaltocfalse}
\bibliography{bib}

\end{document}